\documentclass[aps,twocolumn,showpacs,superscriptaddress,groupedaddress]{revtex4}  
\usepackage{graphicx}  
\usepackage{dcolumn}   
\usepackage{bm}        
\usepackage{amssymb, amsmath, amsfonts, amsopn}   
\usepackage{lipsum}        
\usepackage{color}
\usepackage{ulem}
\usepackage{mathtools}

\DeclareGraphicsExtensions{{.pdf}}
\graphicspath{{./figures/}}

\DeclareMathOperator{\sign}{\mathrm{sign}}
\renewcommand{\vec}[1]{\boldsymbol{#1}}
\newcommand{\uvec}[1]{\hat{\vec{#1}}}
\renewcommand{\tensor}[1]{\mathcal{#1}}
\newcommand{\abs}[1]{\left\lvert{#1}\right\lvert}

\begin{document}

\widetext



\title{
Do hydrodynamically assisted binary collisions lead to orientational ordering of microswimmers?
}
\author{Norihiro Oyama}
\affiliation{Mathematics for Advanced Materials-OIL, AIST-Tohoku University, Sendai 980-8577, Japan}
\author{John Jairo Molina}
\affiliation{Department of Chemical Engineering, Kyoto University, Kyoto 615-8510, Japan}
\author{Ryoichi Yamamoto}
\affiliation{Department of Chemical Engineering, Kyoto University, Kyoto 615-8510, Japan}
\affiliation{Institute of Industrial Science, The University of Tokyo, Tokyo 153-8505, Japan.}
\date{\today}

\begin{abstract}{
We have investigated the onset of collective motion in
systems of model microswimmers, by performing a comprehensive
analysis of the binary collision dynamics using three dimensional direct numerical
simulations~(DNS) with hydrodynamic interactions.
From this data, we have constructed a simplified
binary collision model~(BCM) which accurately reproduces the
collective behavior obtained from the DNS for most cases.
Thus, we show that global alignment can mostly arise solely from
binary collisions.
Although the agreement between both models (DNS and BCM)
is not perfect, the parameter range in which notable differences
appear is also that for which strong density fluctuations
are present in the system
(where pseudo-sound mound can be observed\cite{Oyama2016a}).  }
  \end{abstract}

\maketitle

\section{Introduction}
Active matter encompasses a vast range of systems, from
microorganisms at the microscopic scale, to humans and other mammals
at the macroscopic scale\cite{Vicsek2012,Marchetti2013}.
These active systems can show various nontrivial
collective behaviors
\cite{Lushi2014,Ezhilan2013,Oyama2016a,Chate2008,Bazazi2008,Buhl2006,Zottl2016}.
Especially striking is the global ordering in the absence of
any external field or leading
agents\cite{Ballerini2008,Volfson2008,Lukeman2010,Schaller2010,Evans2011}.
In order to validate the various scenarios that have been proposed to explain
such phenomena, it is important to develop model systems which can be
well-controlled experimentally and efficiently
simulated. For this purpose, micro-swimmers,
such as microbes and self-propelled
Janus particles, have been extensively used.
It is known that hydrodynamic interactions can play a dominant
role in the dynamics of microswimmers dispersions\cite{Rafai2010}.
For example, in ref.~\cite{Lushi2014}, the authors study the dynamics
of confined bacterial suspensions, and found that they could reproduce
the experimentally observed results with simulations, but only if
hydrodynamic interactions were taken into account.
While the experimental
realizations can be rather complicated, simple computational models
exist which allow for direct numerical calculations, such that the
hydrodynamic effects can be accurately represented.
Indeed, several simulation works on model microswimmer dispersions
have recently been
performed in order to study the role of hydrodynamics\cite{Oyama2016a,Evans2011,Kyoya2015,AlarconOseguera2015,Alarcon2013a,Zottl2014,Li2014a,Navarro2015,Matas-Navarro2014a,Ishikawa2014,Ishikawa2010,Giacche2010,Ishikawa2008,ISHIKAWA2007,ISHIKAWA2007a,ISHIKAWA2006,Ishikawa2006a,Molina2013,Sano2016a}.
Of particular interest for our current work is the study by Evans {\it
  et al.}\cite{Evans2011},
who investigated the conditions under which
polar order appears.
  They used a common microswimmer model called the squirmer model,
  which allows one to easily consider different types of swimmers,
  namely, pushers, pullers and neutral swimmers.
They found that the ordering does not depend
strongly on the volume fraction of swimmers,
but rather on the type
and strength of the swimming.
The fact that we observe ordering in very
dilute dispersions suggests that the mechanisms leading to the collective
alignment do not depend on the volume fraction,
and could be
explained by considering only binary collisions.
Thus, in this work we investigate the onset of polar ordering by
performing a detailed analysis of the collision data obtained from
three dimensional~(3D) simulations of swimmer suspensions with hydrodynamic interactions.
First, we have extended the study of Evans' {\it et al.}, by performing
direct numerical simulations (DNS) of bulk suspensions
over a larger set of parameters.
We have confirmed that
the volume fraction dependence
is weak if the volume fraction is small enough
(if it is high, the polar order collapses).
Second, by simulating binary particle collision events with varying
collision geometries, we have
gathered comprehensive information on the changes in swimming
direction that a particle feels when it undergoes a collision.
If we look at changes in the relative angles of two particles after
the collisions,
the results show different tendencies
depending on the type of swimmer.
Pullers tend to exhibit disalignment when the incoming relative
angle is small,
while pushers exhibit disalignment at intermediate values.
Furthermore, using this binary collision data,
we have constructed a simple binary collision model (BCM)
and used it to study the collective alignment of many particle systems
as a function of swimming type.
The BCM successfully reproduced the emergence of the polar order
except for dispersions of intermediate pullers for which a strong
clustering behavior is reported\cite{Oyama2016a,Alarcon2013a}.

\section{Simulation Methods}
\subsection{The Squirmer Model}
In this work, the squirmer model was used to describe the
swimmers\cite{Lighthill1952,Blake1971}.
Squirmers are particles with modified stick boundary
conditions at their surface which are responsible for the
self-propulsion.
The general form is given as an infinite expansion
of both radial and tangential
velocity components, but for simplicity the radial terms are usually
neglected and the infinite sum is truncated to second
order\cite{ISHIKAWA2007}.
For spherical particles, the surface
velocity is given by
\begin{eqnarray}
\vec{u}^s(\theta) = B_1\left(\sin{\theta} + \frac{\alpha}{2}\sin{2\theta}\right)\uvec{\theta}
\label{Sq_2},
\end{eqnarray}
where $\uvec{\theta}$ and $\uvec{r}$~(we use a caret to denote unit
vectors) are the tangential and radial
unit vectors for a given point at the surface of the particle and
$\theta=\cos^{-1}{(\uvec{r}\cdot\uvec{e})}$
is the polar angle (since the system is axisymmetric around the
swimming direction $\uvec{e}$,
the azimuthal angle does not appear).
The steady-state swimming velocity is determined only by the
coefficient of the first mode $B_1$ (the source dipole),
and the ratio of the first two
modes $\alpha=B_2/B_1$ determines the type and strength of the
swimming. 
When $\alpha$ is negative,
the squirmers are pushers, and generate
extensile flow fields, and when it is positive,
they are pullers and generate contractile flow fields. For the special
case when $\alpha = 0$, we refer to the swimmers as a neutral swimmer
which is accompanied by a potential flow.
  The difference in the type of swimmer can be related to the position
  of the propulsion mechanism along the body. A swimmer whose
  propulsion is generated at the back is a pusher
  (e.g. \textit{E. Coli}), one whose propulsion comes from the front
  is a puller (e.g., \textit{Chlamydomonas Reinhardtii}). The former
  will generate an extensile stress, the latter a contractile
  stress. In contrast, for neutral swimmers such as \textit{Volvox}
  the dipole term is dominant ($B_2 \ll B_1$), resulting in a
  symmetric flow field with no vorticity.
  Approximate values for $\alpha$ of real swimmers have been reported
  as $\alpha\approx-1$ for \textit{E. Coli},
  $\alpha\approx 1$ for \textit{Chlamydomonas} and $\alpha\approx 0$ for \textit{Volvox}\cite{Evans2011}.
In what follows, we refer to $\alpha$ as the swimming parameter.

\subsection{Smoothed Profile Method}
In order to solve for the dynamics of squirmers swimming in a
viscous host fluid,
the coupled equations of motion for the fluid and the solid
particles need to be considered.
Particles follow the Newton-Euler equations of motion:
\begin{align}
                \dot{\boldsymbol{R}}_i &= \boldsymbol{V}_i 
                &\dot{\boldsymbol{Q}}_i &= \text{skew}(\boldsymbol{\Omega}_i)\cdot
                \boldsymbol{Q}_i\\
                M_{\text{p}}\dot{\boldsymbol{V}}_i &= \boldsymbol{F}_i^{\text{H}}+\boldsymbol{F}_i^{\text{C}}
                &\boldsymbol{I}_{\text{p}}\cdot\dot{\boldsymbol{\Omega}}_i &= \boldsymbol{N}_i^{\text{H}}\notag
\end{align}
where $i$ is the particle index,
$\boldsymbol{R}_i$ the position, $\boldsymbol{Q}_i$
the orientational matrix, and $\text{skew}(\boldsymbol{\Omega}_i)$ the
skew symmetric matrix of the angular velocity
$\boldsymbol{\Omega}_i$.
The hydrodynamic force
$\boldsymbol{F}_i^{\text{H}}$ and torque 
$\boldsymbol{N}_i^{\text{H}}$ are computed assuming momentum
conservation to guarantee proper coupling between the fluid and the particles.
To prevent particles from overlapping,
we have also included an excluded volume effect
by introducing a repulsive interaction between particles,
$ \boldsymbol{F}_i^{\text{C}} $,
as a truncated Lennard-Jones potential with (36-18) powers.
  Therefore, particles interact with each other both via long-range
  hydrodynamic interactions and the short-range repulsive force.
  We note that the exact form of the short-range repulsive force is not crucial for the dynamics of non-Brownian particles\cite{Dratler1996}.
The time evolution of the fluid flow field is governed by the
Navier-Stokes equation with the incompressible condition:
\begin{align} 
  \boldsymbol{\nabla}\cdot\boldsymbol{u}_{\rm f} &= 0\\
  \rho_{\text{f}} \left( \partial_t + \boldsymbol{u}_{\text{f}}\cdot\nabla\right)\boldsymbol{u}_{\text{f}}
  &= \nabla \cdot \boldsymbol{\sigma}_{\rm f} \\
  \boldsymbol{\sigma}_{\rm f}&=-p\boldsymbol{I}+\eta_{\rm f} \left\{ \nabla\boldsymbol{u}_{\text{f}} + \left( \nabla\boldsymbol{u}_{\text{f}} \right)^t \right\}
\end{align}
where 
$\rho_\text{f}$ is the fluid mass density, $\eta_{\rm f}$ the shear
viscosity, and $\boldsymbol{\sigma}_{\rm f}$ is the Newtonian stress tensor.
To couple these equations efficiently, we have used the Smoothed
Profile Method (SPM),
which enables us to calculate the solid/fluid two-phase dynamics on fixed grids
with hydrodynamic interactions\cite{Nakayama2005,Kim2006,Nakayama2008,Molina2013}.
In the SPM, the sharp interface between the solid and fluid domains is
replaced by a diffuse one with finite width $\xi$,
and the solid phase is represented by a smooth and continuous
profile function $\phi_{\rm p}$.
This profile function takes a value of 1 in the solid domain, and 0 in
the fluid domain.
By introducing the smoothed profile function,
we can define a total velocity field, $\boldsymbol{u}$, which includes both fluid and
particle velocities, and is defined over the entire computational
domain, as:
\begin{align}
  \boldsymbol{u} = \left( 1- \phi \right)\boldsymbol{u}_{\rm f}
  + \phi \boldsymbol{u}_{\rm p},\notag\\
  \phi\boldsymbol{u}_{\rm p} = \sum_i\phi_i\left[ \boldsymbol{V}_i+
    \boldsymbol{\Omega}_i\times\boldsymbol{R}_i\right],
\end{align}
where, $\left( 1- \phi \right)\boldsymbol{u}_{\rm f}$ is the
contribution from the fluid, $\phi \boldsymbol{u}_{\rm p}$ from
the particle motion.
The time evolution of the total flow field $\boldsymbol{u}$ obeys:
\begin{align}
  \boldsymbol{\nabla}\cdot\boldsymbol{u} &= 0,\notag\\
  \rho_{\text{f}} \left( \partial_t + \boldsymbol{u}\cdot\nabla\right)\boldsymbol{u}
  &= \nabla \cdot \boldsymbol{\sigma}_{\rm f}+\rho_{\rm f}\left(
  \phi\boldsymbol{f}_{\rm p}+\boldsymbol{f}_{\rm sq}\right)
\end{align}
where $\phi\boldsymbol{f}_{\text{p}}$ is the body force necessary
to maintain the rigidity of particles, and
$\boldsymbol{f}_{\text{sq}}$ is the force due to the
active squirming motion.
This method drastically reduces
the computational cost.

\section{Results}
\subsection{Bulk Polar Order}
\begin{figure}
  \begin{center}
  \includegraphics[width=\linewidth]{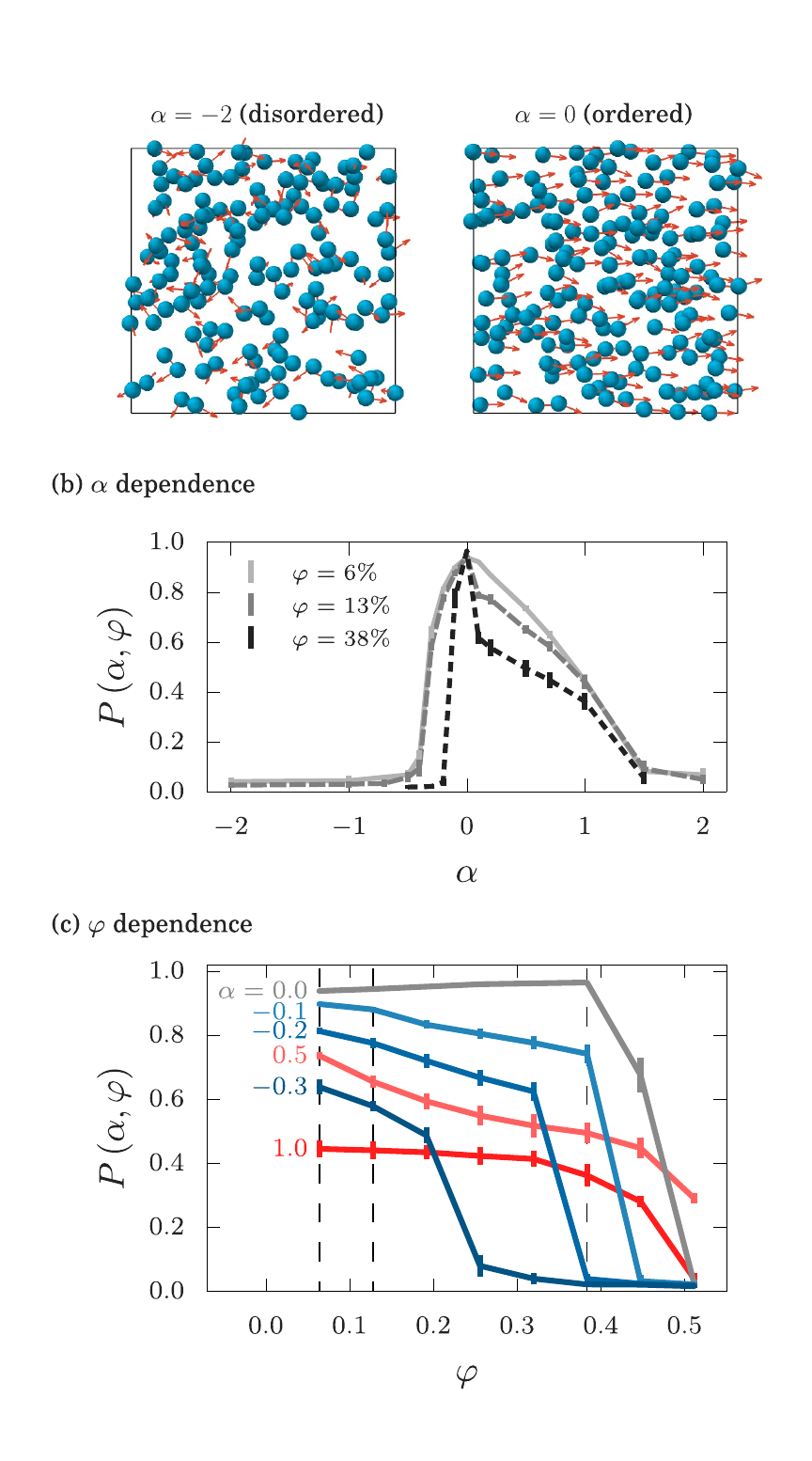}
   \caption{\label{fig:bulkorder}
     (a) Simulation snapshots for disordered~($\alpha=-2$) and
     ordered~($\alpha=0$) states. The arrows give the direction of
     motion, and only a subset of the particles have been drawn.
     (b) The $\alpha$ dependency of the polar
     order $P\left( \alpha, \varphi \right)$ for
     $\varphi=6\%$ (solid line), $13\%$ (light dashed line) and $38\%$
     (dark dashed line).
     (c) The $\varphi$ dependency of $P\left( \alpha, \varphi \right)$
     for several values of $\alpha$. Dashed vertical lines indicate
     the volume fractions of $6,13$ and $38\%$ used in (a) and (b).
   }
   \end{center}
\end{figure}

To start, we used the SPM to carry out DNS studies of bulk
swimmer dispersion in 3D at varying volume fraction $\varphi$ and
swimming type $\alpha$. 
We use a cubic system of linear dimension of $64\Delta$, with $\Delta$
the grid spacing. The viscosity and
the mass density of the host fluid, $\mu$, $\rho_f$ are set to one,
such that the unit of time is $t_0=\rho_f\Delta^2/\mu$.
The particle diameter $\sigma$ and the interface thickness $\xi$ are
$4\Delta$ and $2\Delta$ respectively.
We used a random initial configuration for the particle positions and orientations,
and varied the number of particles $N_{\rm p}$ from $500$ to $4000$, which
correspond to a range of volume fraction $0.06 \lesssim \varphi \lesssim 0.5$.
To quantify the degree of collective alignment,
we calculate the polar order parameter $P$\cite{Evans2011,Alarcon2013a}
\begin{align}
  P =\left\langle \frac{1}{N_{\rm p}} \left|\sum_i^{N_{\rm p}} \uvec{e}_i\right|\right\rangle,
  \label{eq:bulk_order}
\end{align}
where $\uvec{e}_i$ is the swimming direction of particle
$i$, $N_{\rm p}$ the number of particles
and angular brackets denote an average over time (after steady
state has been reached).
Typical simulation snapshots for disordered ($P\simeq 0$) and ordered
($P\simeq 1$) systems are given in Fig.~\ref{fig:bulkorder}(a).
We note that even for a completely random distribution of
orientations, the polar order defined by
Eq.~(\ref{eq:bulk_order}) will not be exactly $0$, and will depend
slightly on the number of particles, as $P_0=1/\sqrt{N_{\rm p}}$,
where $P_0$ represents the polar order value below which we can
consider the system is in an isotropic phase.

The polar order parameter $P$ is a function of only the volume
fraction of particles $\varphi$ and the swimming parameter $\alpha$\cite{Evans2011}.
First, we investigated $\alpha$ dependency.
The results are illustrated in Fig.~\ref{fig:bulkorder}(b), for volume
fractions $\varphi=6\%$ ($N_{\rm p} = 500$),$13\%$ ($N_{\rm p} =
1000$) and $38\%$ ($N_{\rm p} = 3000$).
All the results show a similar tendency and are
in agreement with preceding works\cite{Evans2011,Alarcon2013a}:
$P$ has a maximum at $\alpha=0$, independent of $\varphi$,
and decreases with increasing value of $\abs{\alpha}$.
In addition, the $P$ for pushers decays faster than that
of pullers as the magnitude of $\alpha$ is increased.
For non-pusher $\alpha \ge 0$, the volume fraction dependence of $P$ is not very
large and at least the qualitative ordering tendency is
the same; but for weak pushers (for example $\alpha \approx -0.3$),
we observe a significant drop in the value of $P$, or an
order/disorder phase transition when the volume fraction increases.
The volume fraction dependence can be seen clearly in
Fig.~(\ref{fig:bulkorder}c), where we have plotted the values of $P$ over the entire
volume fraction range for six different swimmers ($\alpha = -0.3, -0.2, -0.1, 0, 0.5, 1$).
Evans {\it et al.} have previously reported such a volume fraction dependence
for $\alpha=1$~\cite{Evans2011}.
To understand the dependence of $P$ on the swimming type $\alpha$,
in particular the different behaviors seen for pushers and pullers,
it is useful to compare them against the results obtained for neutral
swimmers $\alpha=0$, which show the highest degree of alignment.
As seen in Figure\ref{fig:bulkorder}~(c), the order parameter for
$\alpha=0$ shows two distinct regimes: for $\varphi\lesssim 0.4$ there
is little variation; for $\varphi \gtrsim 0.4$ there is
a drastic drop in the order parameter to $P=0~(P_0)$.
The same behavior is observed for pushers,
although both of the degree of ordering and the critical volume fraction $\varphi_c$
(where the order parameter falls to zero) are both reduced (higher
$\abs{\alpha}$ resulting in lower $P$ and $\varphi_c$).
In contrast, pullers show a gradual decrease only in the degree of
order depending on $\left|\alpha\right|$.
Interestingly, intermediate pullers ($\alpha= 0.5$) 
maintain a non-zero order parameter
over the entire volume fraction range we have considered
(all other systems giving $P\approx P_0$ at the highest $\varphi$).
We believe this anomalous behavior for the intermediate
pullers can be related to the strong clustering behavior that gives rise to density
inhomogeneities\cite{Oyama2016a,Alarcon2013a}.
Note that because the number of particles are sufficiently large
(even for the system with the smallest volume
fraction which corresponds to $N_{\rm p}=500$,
the value of $P_0$ is less than 0.05),
the decays
in $P$ are not related to the fact that we use different
values of $N_{\rm p}$.
  We note that continuum theories predict an unstable long wave-length
  ordering, with no global order in the limit of infinitely large
  systems\cite{AditiSimha2002}. However, we consider that the finite
  size effects, if they exist, will not lead to qualitatively
  different results. This is supported by the fact that
  the pair distribution function decays very fast for most squirmer
  dispersions (see Supplemental Material). Only in the case of
  pullers do we measure a long-range correlation which might suffer
  from finite size effects. In fact, these effects were studied in detail by
  Alarc\'{o}n\cite{AlarconOseguera2015}, who nevertheless
  showed that the polar order converges to a non-zero value as the
  system size is increased. The discrepancies with the continuum
  predictions are likely due to the absence of the finite particle
  volume term in such theories, which only take into account the
  long-range hydrodynamic interactions. While these long-ranged
  interactions tend to destabilize the global ordering, in squirmer
  dispersions they are screened by neighboring particles, allowing
  the system to maintain its order, even for very large systems.

\begin{figure}
  \begin{center}
  \includegraphics[width=0.9\linewidth]{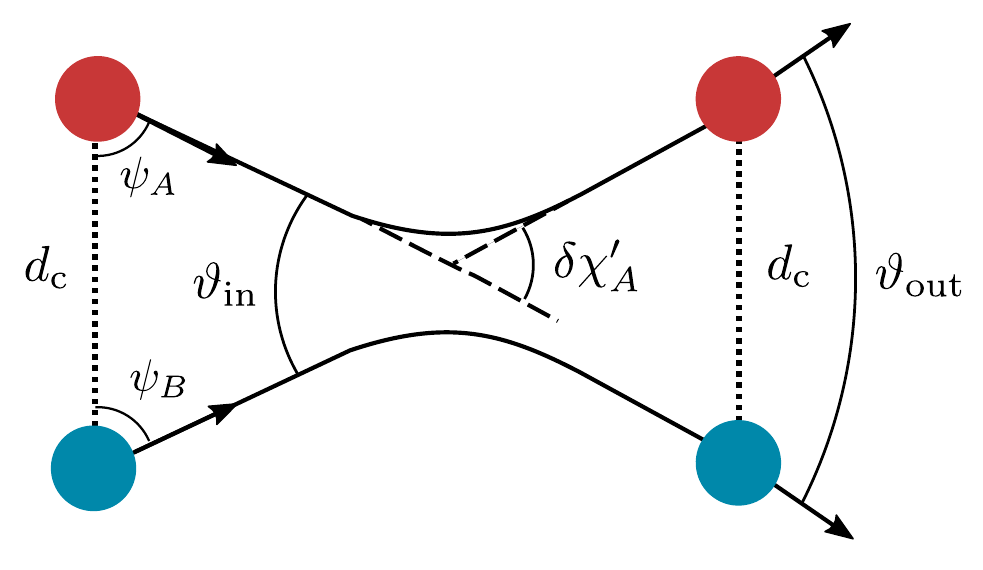}  
   \caption{\label{fig:theta} Schematic representation of the
     collision geometry.
   }
   \end{center}
\end{figure}

\begin{figure*}
  \begin{center}
  \includegraphics[width=0.8\textwidth]{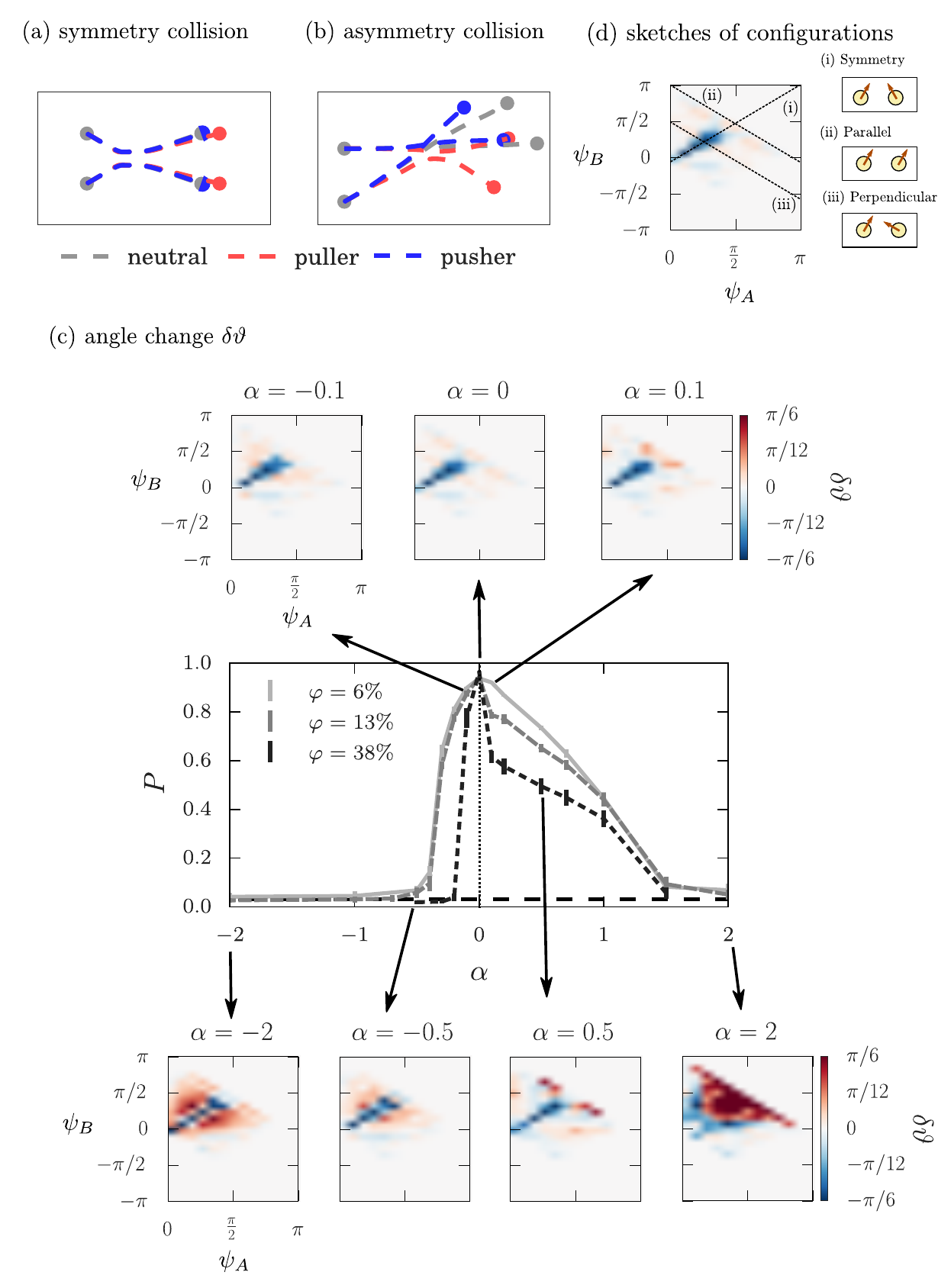}  
  \caption{\label{fig:scat}
    (a-b) Typical trajectories of collisions of neutral
    swimmer~($\alpha=0$; colored
    as gray), pullers~($\alpha=0.5$; red) and
    pushers~($\alpha=-0.5$; blue) for (a)~a symmetric collision
    and (b)~an asymmetric one.
    (c) Change in the relative angle between the particles during collisions,
    $\delta\vartheta=\vartheta_{\rm out}-\vartheta_{\rm in}$, as a
    function of the initial orientations, $\psi_{\rm A}$
    and $\psi_{\rm B}$.
    Results of bulk polar order measurement are also shown~($\varphi=6,13$ and $38\%$).
    In intensity maps, red colors mean positive values or the disalignment effect and
    blue colors negative or the
    aligning effect.
    (d) Sketches for the characteristic configurations in the intensity maps in (c).
   }
   \end{center}
\end{figure*}

\subsection{Binary Collision Analysis}\label{sec:BCM}
Taking into account the fact that at low volume fractions,
two body interactions are dominant,
we can expect that the observed polar order in bulk is due to binary collisions.
This is supported by the fast decay in the spatial correlations of
the particle velocities. In the Supplemental Material, we show
results for the velocity correlations of systems at $\varphi=0.06$
for both squirmers and inert sedimenting colloids.
For the squirmers, regardless of the swimming parameter $\alpha$, the
correlation is nonzero only in the close vicinity of the particle,
while the correlation length for the colloidal systems extends to several particles diameters.
Such short-ranged correlations for swimmers at low volume fractions suggest that only binary collisions can lead to
the polar order observed in bulk.
To verify the hypothesis proposed above, we first conducted an intensive analysis on the
binary collision of squirmers with varying values of $\alpha$.
For this, we conducted simulations with only two squirmers in a quasi two-dimensional setup,
where particles are confined to a 2D plane, while the computational domain is fully three dimensional.
Then, we tried to construct a simplified binary
collision model~(BCM) using the data obtained by this analysis.
We note that a similar binary collision analysis for pullers has been done by Ishikawa
{\it et al.}\cite{ISHIKAWA2006}.
We have extended their work to pushers and neutral swimmers and made direct comparison
between the BCM and the bulk DNS results.

We have carried out 3D DNS
for a pair of particles with various collision geometries
and $\alpha$ values.
Given the symmetry of the problem, the two
particles will move in a 2D plane (defined by the two orientation
vectors).
We considered collisions of two particles labeled $A$ and $B$.
The precise parametrization we have used to describe the collision is
given in Fig.~\ref{fig:theta},
where three sets of angles have been
defined, $\psi_j$, $\delta\chi_j$ ($j\in\left\{ A,B\right\}$ is the particle label) and
$\vartheta_{\rm in/out}$.
The initial configuration of the system is specified by $\psi_j$,
the angles between the direction
of motion and the center-to-center distance vector at the initial state.
These angles determine whether particles start swimming towards or away
from each other.
The information for the change in the 
swimming direction of each particle is given by $\delta\chi_{j}$.
Then, the relative orientation of particles when the collision event
starts/ends is represented by $\vartheta_{\rm in/out}$.
Due to the long-range nature of the hydrodynamic interactions,
  particles can alter their directions even without touching,
  in contrast to collisions in a gas of hard-sphere particles.
  Therefore, there is no unique way to define a ``collision''
  between particles.
  In this work, we define a characteristic distance $d_c$ that
  is the threshold distance under which particles are considered
  to be colliding: a collision event has started when the distance
  between the two particles becomes less than $d_c$, and it lasts until
  the distance exceeds this value (see Figure~\ref{fig:theta}).
Thus, $d_c$ should be large enough
that hydrodynamic interactions can be
neglected when the distance between the particles exceeds $d_c$.

The parameters for
the binary collision is determined as follows.
The initial particle distance was set to $d_0 = 16\Delta=4 \sigma$, and the
collision threshold to $d_c=15\Delta$.
The value of $d_c$ is determined to be big enough so that we can
safely ignore the hydrodynamic interactions if the particle-particle
distance is greater than $d_c$
(above this value, particles hardly change their orientations).
The value of $d_0$ is determined so that swimmers have obtained their
steady state velocity when the inter-particle distance becomes $d_c$.
The initial geometry was varied by changing $\psi_j$ in intervals of
$\pi/12$, for $0 \le \psi_A \le \pi$ and $-\pi \le \psi_B \le \pi$.
To take into account the symmetry of the system,
we label one of the particles ($A$) as a reference
particle, and take $\psi_A \ge 0$, while $\psi_B$ is defined as
\begin{align}
  \psi_B &=
  \sign{\left(\tensor{P}_{AB}\cdot\uvec{e}_A\right)}\sign{\left(\tensor{P}_{AB}\cdot\uvec{e}_B
    \right)} \abs{\arccos{\left(\uvec{r}_{AB}\cdot\uvec{e}_B\right)}}
\end{align}
where $\vec{r}_{AB} = \vec{r}_B - \vec{r}_A$, $\tensor{P}_{AB}$ is the
projection operator (with $\tensor{I}$~the~identity~operator)
\begin{align}
  \tensor{P}_{AB} &= \tensor{I} - \uvec{r}_{AB}\uvec{r}_{AB}
\end{align}
and $\sign(x) = 1$ for $x\ge 0$ and $\sign(x) = -1$ for $x< 0$.
As mentioned above, we use a caret to denote unit vectors.
Thus, $\psi_B$ is defined as positive if both particles are swimming towards
the same side with respect to the center-to-center line between particles.
We note that only combinations of $\psi_j$ which meet
$\hat{\boldsymbol{e}}_{AB}\cdot\hat{\boldsymbol{r}}_{AB}<0$ can lead to
``collisions'', where $\boldsymbol{e}_{AB}=\hat{\boldsymbol{e}}_B-\hat{\boldsymbol{e}}_A$.
Three-dimensional simulations for the
binary collision were performed using the same system parameters as
for the bulk simulations presented above.

The results of the binary collision analysis are summarized in
Fig.~\ref{fig:scat}.
Fig.~\ref{fig:scat}(a) and (b) show typical trajectories and
Fig.~\ref{fig:scat}(c) shows
changes in the relative angle
between the swimming direction of the two particles after
the collisions, $\delta
\vartheta=\vartheta_{\rm out}-\vartheta_{\rm in}$.
In the following, we define a symmetric collision
as a collision in which
$\psi_A=\psi_B$~(Fig.~\ref{fig:scat}(a)).
Intensity maps show the values of $\delta\vartheta$ as a function of
the initial angles, $\psi_j$.
In Fig.~\ref{fig:scat}(d),
the schematic representations of three characteristic
initial configurations are shown: (i)~symmetric, (ii)~parallel, (iii)~perpendicular.
Although the parallel configuration does not lead to a collision,
it is useful to identify the corresponding region in the intensity
plots shown in Fig.~\ref{fig:scat}(c).
The values of the polar order in bulk are again shown to make
the connection between the bulk and binary collision dynamics clear.
As shown in Fig.~\ref{fig:scat}~(a) and (b),
different values of $\alpha$ lead to different 
particle trajectories,
resulting in different patterns for
$\delta\vartheta$~(Fig.~\ref{fig:scat}~(c)).
The results of $\delta\vartheta$ in systems of pushers and pullers 
can be easily understood by considering their deviation from
the results for neutral swimmers~($\alpha=0$).
The neutral swimmers
show strong aligning behaviors only when the collision is symmetric,
and just small absolute values of $\delta \vartheta$ otherwise.
If we look at the results for pullers~($\alpha>0$),
we can perceive that, in the case of $\alpha=0.1$,
disalignment effects are detected at 
small relative incoming angles.
Such disalignment effects 
becomes stronger with the increase in the absolute value of $\alpha$,
as shown in the subplots for $\alpha=0.5,2$.
On the other hand, in the cases of pushers~($\alpha<0$),
disalignment effects are seen at relatively large incoming angles.
For pushers, as well as for pullers,
the increase in the absolute value of
$\alpha$ leads to stronger disalignment effect.
In this way, measuring only $\delta\vartheta$,
we can observe different tendencies between pushers and pullers.
These tendencies seem to be a consequence of the complicated hydrodynamic interactions,
and it is impossible to understand intuitively from
the view point of the flow field which a single swimmer generates.

To implement a simple binary collision model using the collision data
obtained from the DNS,
it is necessary to measure the
changes in the single particle orientations, $\delta\chi_j$.
For this,
from the comprehensive DNS data for binary collisions,
we have determined $\delta\chi_j$ for all the collisions, as
\begin{align}
  \delta\chi_j =
  \arcsin\left(\uvec{z}_\chi\cdot\left( \uvec{e}_j^{\rm
    in}\times\uvec{e}_j^{\rm out}\right)\right),\label{eq:delchi}
\end{align}
  where the superscript ``in/out'' refers to the value at the moment when a
collision starts/ends,
$\uvec{z}_\chi=\frac{\uvec{e}_{j}^{\rm in}\times\uvec{e}_{j'}^{\rm in}}{|
  \uvec{e}_{j}^{\rm in}\times\uvec{e}_{j'}^{\rm in}|}$ and $j'$ refers
to the particle which is colliding with particle $j$.

\begin{figure}
  \begin{center}
  \includegraphics[width=\linewidth]{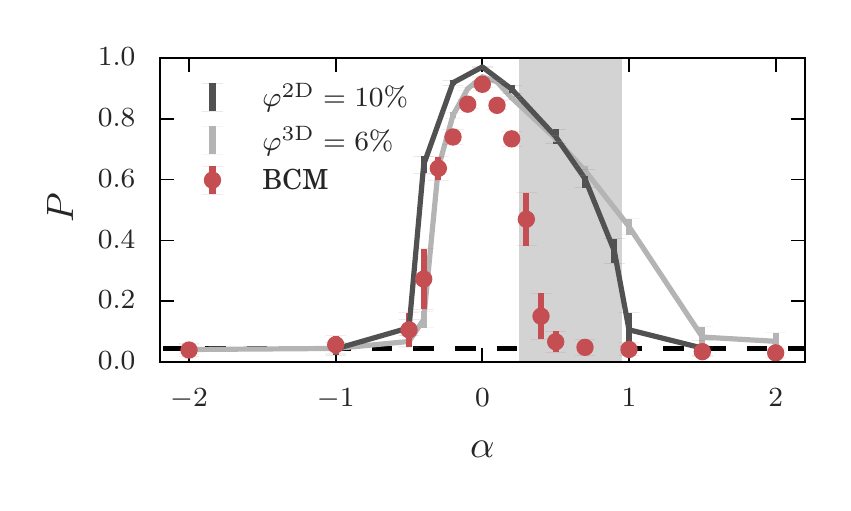}
   \caption{\label{fig:quasi2D}
     The $\alpha$ dependency of the polar
     order $P\left( \alpha, \varphi \right)$ for
     3D bulk system with $\varphi^{\rm 3D}=6\%$ (light line), quasi-2D bulk system with $\varphi^{\rm 2D}=10\%$ (dark line).
     Results for the simplified binary collision model are given as circles.
   }
   \end{center}
\end{figure}

\begin{figure*}
  \begin{center}
  \includegraphics[width=\linewidth]{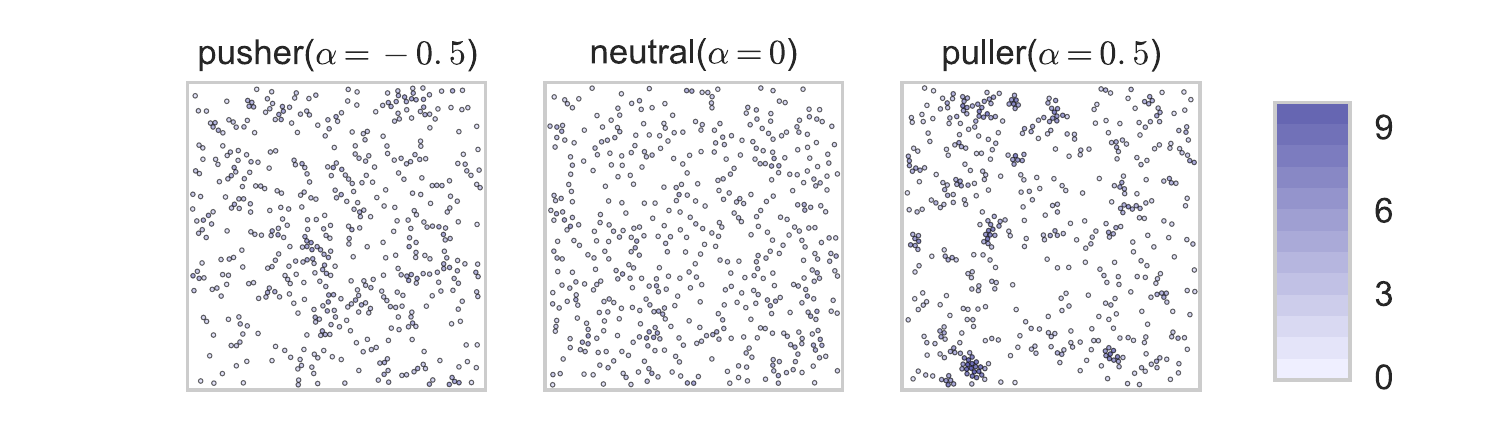}
   \caption{\label{fig:snap}
     Typical snapshots in quasi-2D bulk systems.
     The results for the systems with $\alpha=0, \pm0.5$ are shown.
     The color represents the number of particles with in the distance of $2\sigma$.
   }
   \end{center}
\end{figure*}

Finally, in order to investigate whether the polar order seen in bulk systems can be
explained only by binary collisions,
we constructed a binary collision model~(BCM).
Here, we have necessarily introduced two simplifications.
First, we assume 2D systems.
And second, we consider only binary collisions, and use the
statistics of collision angles obtained from the present DNS.
Because we are assuming very dilute system
such that the information of the position doesn't matter anymore,
the particles have only the information about the orientations.
Under these simplifications, we calculated the polar order of the
system of BCM using the following simple algorithm.
At each step of the simulation,
we randomly choose two particles~(let's say particles $i$ and $i^\prime$).
The selected particles will experience a ``collision'',
which will change  their orientations according to the statistics
obtained from the binary collision analysis:
\begin{align}
  \chi_i \left( s + 1 \right)&=\chi_i\left( s \right)+
  \delta\chi_i,\notag\\
  \chi_{i^\prime} \left( s + 1 \right)&=\chi_{i^\prime}\left( s \right)+
  \delta\chi_{i^\prime},\\
  \chi_k \left( s + 1 \right)&=\chi_k\left( s \right)\notag,
\end{align}
where subscript $k$ stands for the particles which are not selected
to collide.
The values $\delta\chi_{i}$ and $\delta\chi_{i^\prime}$
are random numbers generated according to the conditional probability
distribution when the relative incoming angle
$\vartheta_{\rm in}$ is given:
$P\left(\delta\chi_i, \delta\chi_{i^\prime} | \vartheta_{\rm
  in}\left(i,i^\prime\right)\right)$,
where $\delta\vartheta\left(i,i^\prime \right)$ means the relative
incoming angle between particles $i$ and $i^\prime$.
The conditional probability distribution $P\left(\delta\chi_i, \delta\chi_{i^\prime} | \vartheta_{\rm
  in}\left(i,i^\prime\right)\right)$ is determined by using the
results of the binary collision analysis presented above.
Because there is the information about only the
orientation in the BCM,
the orientation update algorithm is based only on the relative incoming angle
$\bar{\vartheta}_{\rm in}$,
and does not depend on the collision parameter (which cannot be
defined in this model system) or other geometrical
information.
No noise term is included.
After a sufficiently large number of collisions,
the system reaches a steady state, with a constant polar order.
  We conducted calculations using this BCM for various values of $\alpha$,
  while keeping the value of $N_{\rm p} = 500$ constant. 
  The results for these simulations are plotted in Fig.~(\ref{fig:quasi2D}) as red circles, 
  together with results for the quasi-2D and 3D bulk DNS (dark and
  light solid lines, respectively). The quasi-2D bulk simulations were
  included for a fair comparison with the BCM
  results, since the latter is itself obtained from quasi-2D DNS.
  The setup for these quasi-2D bulk simulations is as follows. 
  The computational domain is three-dimensional, with linear
  dimensions $L_x = 64\sigma, L_y = 64\sigma$ and $L_z = 4\sigma$
  under full periodic boundary conditions. The remaining parameters
  are the same as those for the 3D bulk systems. 
  Particles are initially placed within the $x-y$ plane at $z =
  L_z/2$, which we refer to as the center plane. 
  The particles are allowed to rotate only around the $z$-axis, such that
  their trajectories are confined to this center plane. 
  The number of particles is 500 (the same value used in the BCM calculations), 
  which corresponds to an area fraction of $\varphi^{\rm 2D}\approx 10\%$.
  We use $\varphi^{\rm 3D}$ for the volume fraction in 3D system and $\varphi^{\rm 2D}$ for the area fraction in quasi-2D system. 
  The BCM results and those from the quasi-2D bulk DNS
  are in good agreement with each other for non-pullers ($\alpha \le 0$).
  Interestingly, the results from the 3D bulk DNS also fit very well those of BCM and the quasi-2D bulk DNS.
  This implies that the dimensionality does not play a big role in determining the polar order formation in squirmer dispersions.
This also indicates that the appearance of polar order
can be understood just in terms of binary collision events for non-pullers.
For pullers, we see an increasing deviation:
the larger $\alpha$ becomes, the larger the deviation becomes.
For $\alpha\ge 0.4$, qualitatively different results are
obtained: in particular, for the BCM the order has collapsed.
  Here, let us consider the cause of the discrepancy.
  The BCM is missing two main aspects which 
  affect the dynamics of swimmers: namely, 
  correlated collisions and many body nature of the hydrodynamic interactions.
  First, in the BCM, the correlation between collisions and
  particle positions is neglected and the system dynamics is
  determined by repeated uncorrelated collisions in which
  the absolute and relative outgoing angles are drawn from
  the probability distribution measured from the binary
  system DNS.
  In real dense dispersions, on the contrary,
  a sequence of collisions which one particle experiences
  can be correlated.
  Second, the BCM assumes that the interactions
  can be considered as a superposition of binary collisions and
  therefore ignores the many body nature of 
  the hydrodynamic interactions, which couples the dynamics of particles in real dispersions.
  Both these effects are expected to become non-negligible and lead to changes
  in the probability distribution of outgoing angles when the local density is high.
  The discrepancy between bulk DNS and the BCM can be understood
  as indirect evidence for the importance of these multi-particle interactions
  on the order formation in the case of intermediate pullers.
  The shaded gray region in Fig.~\ref{fig:quasi2D} marks the parameter range
  in which we have observed strong clustering in bulk systems\cite{Oyama2016a};
  indeed, it is precisely
  in this region where the results do not coincide with the
  BCM (in Fig..~\ref{fig:snap}, typical snapshots for the systems with
  $\alpha =0,\pm 0.5$ in quasi-2D bulk system are shown).
Though several efforts have been dedicated to verify the importance of
binary collisions to explain the polar order formation for various
systems both experimentally and numerically
\cite{Katz2011a,Hanke2013a,Lam2015a,Suzuki2015a,Hiraoka2016},
the presented work is the first successful attempt to conduct such
analysis considering full hydrodynamics.

\section{Conclusion}
Using DNS for squimer dispersions, we have investigated the emergence
of polar ordering and its dependency on the particle volume
fraction $\varphi$ and swimming strength $\alpha$.
In agreement with a previous work\cite{Evans2011}, we
see that the volume fraction dependence is rather weak, and the
ordering depends mostly on $\alpha$ when $\varphi$ is small enough,
while at a large value of volume fraction, we observe an
order/disorder phase transition.
Still, we observed novel volume fraction dependencies for
$\abs{\alpha}<1$.
In particular,
intermediate pullers show no decay of the polar order even at a very high volume
fraction, at which all other swimmers show a decay.
We believe this anomalous behavior at such a high volume fraction
reflects the already-known strong clustering characteristics\cite{Oyama2016a,Alarcon2013a}.
On the other hand, weak pushers show a decay of the polar order even at
small volume fractions.

We conducted a detailed analysis of the
binary collision dynamics of two swimmers and looked at the changes
in the relative orientation of two swimmers after the collisions.
The results show different qualitative disaligning tendencies between pusher
and puller: pullers show disaligning effects at small relative
incoming angles while pushers exhibit at relatively large angles.
The absolute value of $\alpha$ changes only the magnitude of
disalignment, and the tendency is determined by the sign.
Such an analysis also enabled us to construct a
simple binary collision model which is able to reproduce the polar ordering
seen in the bulk DNS for pushers and neutral swimmers.
Thus, it seems binary collisions are
enough to explain the appearance of long range polar ordering for
these types of swimmers.
We note that intermediate pullers exhibit a clear discrepancy between the DNS
results and the BCM;
however, this occurs in the parameter range
where strong clustering behavior is also observed.
This can be seen as indirect evidence that in intermediate puller systems,
multi-body interactions play an important role.
In other words, the origin of the polar order formation can be
different, depending on the specific type of swimming.
In particular, the mechanism responsible for the clustering of intermediate
pullers is still an open question.

\section{Acknowledgement}
We thank N. Yoshinaga, M. Tarama, H. Ito, K. Ishimoto and Simon K. Schnyder for enlightening discussions.
This work was supported by the Japan Society for the Promotion of
Science (JSPS)
KAKENHI Grant No.~17H01083 and also by a Grant-in-Aid for
Scientific Research on Innovative Areas
``Dynamical ordering of biomolecular systems for creation of
integrated functions''(No.~16H00765)
from the Ministry of Education, Culture, Sports, Science, and Technology
of Japan. 

\bibliographystyle{apsrev4-1}
%

\end{document}